\documentclass[12pt,a4]{article}
\pdfoutput=1
\usepackage{amsmath,amssymb}

\newcommand{\im}{\mathrm i}

\newcommand{\tr}{\operatorname{Tr}}

\newcommand{\ket}[1]{\left|#1\right\rangle}      
\newcommand{\eq}{\begin{equation}}
\newcommand{\en}{\end{equation}}
\newcommand{\bear}{\begin{eqnarray}}
\newcommand{\ear}{\end{eqnarray}}

\title{Bethe ansatz for the Temperley-Lieb spin-chain with integrable open boundaries}

\author{G.A.P. Ribeiro\footnote{pavan@df.ufscar.br}, A. Lima-Santos\footnote{dals@df.ufscar.br} \\  Departamento de F\'{i}sica, Universidade Federal de S\~ao Carlos \\ 13565-905 S\~ao Carlos-SP, Brazil}

\begin{document}
\maketitle

\begin{abstract}
In this paper we study the spectrum of the spin-$1$ Temperley-Lieb spin chain with integrable open boundary conditions. We obtain the eigenvalue expressions as well as its associated Bethe ansatz equations by means of the coordinate Bethe ansatz. These equations provide the complete description of the spectrum of the model.
\end{abstract}
\centerline{Keywords: Temperley-Lieb model, Coordinate Bethe ansatz, K-matrices}

\thispagestyle{empty}

\newpage

\newpage

\section{Introduction}

Quantum integrable models and their associated classical vertex models have been largely investigated over the years \cite{BAXTER,KOREPIN}. Many of these models have been widely studied by Bethe ansatz techniques with different boundary conditions and at zero or finite temperature and magnetic field.

Nevertheless one still has interesting problems withstanding to the standard techniques. One of these problems is the biquadratic spin-$1$ model \cite{PARKI,KLUMPER90,ALCARAZ}. The biquadratic model was shown to be invariant by the Temperley-Lieb algebra \cite{BATCHELOR3}. This property has made possible the discovery of new integrable quantum spin chains, which was achieved by exploiting the representation theory of the Temperley-Lieb algebra \cite{BATCHELOR1,BATCHELOR2}.

Moreover the biquadratic model and its generalizations (Temperley-Lieb spin chains) were solved by coordinate Bethe ansatz for periodic boundary conditions and free ends \cite{PARKI,LIMA1,LIMA2,GHIOTTO}. Afterwards, it has also appeared the spectrum of the transfer matrices by means of functional methods\cite{KULISH1}. However, there is still no algebraic Bethe ansatz formulation for these models.

Recently the concept of Temperley-Lieb equivalence\cite{BAXTER} was used in order to obtain the spectral properties of quantum spin chains of Temperley-Lieb type for periodic boundary conditions and free ends \cite{KLUMPER}. The study of the spectral multiplicities allowed to the computation of the thermodynamic properties at finite temperature \cite{KLUMPER}. 

Besides that, the solution of reflection equation, associated to the problem of Temperley-Lieb spin chains with integrable open boundaries, was recently obtained \cite{LIMA3,KULISH2}. However, the computation of the spectra of these spin chains is still an open problem.

In order to fill this gap, in this paper we are interested in the spectra of the $U_q\left[sl(2)\right]$ Temperley-Lieb spin chain with integrable open boundary conditions \cite{LIMA3}. We use a suitable generalization of the coordinate Bethe ansatz\cite{LIMA2} in order to obtain the eigenvalues of the spin-$1$ Temperley-Lieb model with diagonal open boundaries.

The outline of the article is as follows. In section \ref{basics} we introduce the Temperley-Lieb spin chain with integrable open boundaries. In section \ref{cBA} we discuss the application of the coordinate Bethe ansatz and obtain the eigenvalues of the Temperley-Lieb spin chain. Our conclusions are given in the section \ref{CONCLUSION}.

\section{Temperley-Lieb spin chain}
\label{basics}

The cornerstone of the theory of quantum integrable models in one-dimension is given by the Yang-Baxter equation, 
\eq
\check{R}_{12}(\lambda-\mu)\check{R}_{23}(\lambda) \check{R}_{12}(\mu) =\check{R}_{23}(\mu)\check{R}_{12}(\lambda) \check{R}_{23}(\lambda-\mu).
\label{yangbaxter}
\en
This equation provides the commutativity property of the transfer matrix $T(\lambda)=\tr_{\cal A}\left[{\cal T}_{\cal A}(\lambda) \right]$, where ${\cal T}_{\cal A}(\lambda)=R_{{\cal A}L}(\lambda)\cdots R_{{\cal A}1}(\lambda)$, $R_{12}(\lambda)=P_{12}\check{R}_{12}(\lambda)$ and $P_{12}$ is the permutation operator. Quantum integrable spin chain with periodic boundary conditions are obtained by means of the logarithmic derivative of the transfer matrix $T(\lambda)$.

The Temperley-Lieb invariant solutions of the Yang-Baxter equation (\ref{yangbaxter}) are well known\cite{MARTIN}. The spin-$1$ $U_q\left[sl(2)\right]$ solution can be written as,
\eq
\check{R}_{ij}(\lambda)=\frac{\sinh(\gamma-\lambda)}{\sinh{\gamma}} I_{ij} + \frac{\sinh{\lambda}}{\sinh{\gamma}} U_{ij},
\label{Rmatrix}
\en
where $2\cosh{\gamma}=q^2+1+q^{-2}$ and the Temperley-Lieb operator is given by
\eq
U_{12}=\left(\begin{array}{ccc|ccc|ccc}
       0 & 0 & 0 & 0 & 0 & 0 & 0 & 0 & 0 \\
       0 & 0 & 0 & 0 & 0 & 0 & 0 & 0 & 0 \\
       0 & 0 & q^{-2} & 0 & -q^{-1} & 0 & 1 & 0 & 0 \\
\hline
       0 & 0 & 0 & 0 & 0 & 0 & 0 & 0 & 0 \\
       0 & 0 & -q^{-1} & 0 & 1 & 0 & -q & 0 & 0 \\
       0 & 0 & 0 & 0 & 0 & 0 & 0 & 0 & 0 \\
\hline
       0 & 0 & 1 & 0 & -q & 0 & q^2 & 0 & 0 \\
       0 & 0 & 0 & 0 & 0 & 0 & 0 & 0 & 0 \\
       0 & 0 & 0 & 0 & 0 & 0 & 0 & 0 & 0 \end{array} \right),
\en
which is the projector onto the two-sites spin zero singlet written in the basis $\{\ket{+},\ket{0},\ket{-} \}$.

The notion of integrability was extended to tackle with open boundary problems \cite{SKLYANIN}. On the one hand, the $R$-matrix describes the bulk dynamics, and on the other hand a new set of matrices, the $K$-matrices, represent the interactions at the left and right ends of the open spin chain. This is a consequence of the reflection equation, which reads
\eq
R_{12}(\lambda-\mu)K_{1}(\lambda)R_{21}(\lambda+\mu)K_{2}(\mu) =K_{2}(\mu) R_{12}(\lambda+\mu) K_{1}(\lambda) R_{21}(\lambda-\mu).
\label{refle}
\en

In the case of open boundary conditions, the transfer matrix can be written as
\eq
t(\lambda)=\tr_{\cal A}\left[K_{\cal A}^{(+)}(\lambda){\cal T}_{\cal A}(\lambda)K_{\cal A}^{(-)}(\lambda)\left[{\cal T}_{\cal A}(-\lambda)\right]^{-1} \right],
\label{openT}
\en
where $K_{\cal A}^{(-)}(\lambda)$ can be chosen as one of the solutions of the reflection equation (\ref{refle}). The other boundary matrix $K_{\cal A}^{(+)}(\lambda)$ is obtained from the previous one by means of the isomorphism \cite{NEPO},
\eq
K_{\cal A}^{(+)}(\lambda)=K_{\cal A}^{(-)}(-\lambda-\rho)^t V^{t} V,
\label{isomorphism}
\en
where $t$ means transposition, the crossing parameter is $\rho=-\gamma$ and the crossing matrix is given by
\eq
V=\left(\begin{array}{ccc}
       0 & 0 & q \\
       0 & -1 & 0 \\
       q^{-1} & 0 & 0 
\end{array} \right).
\en

The integrable open spin chain is obtained by means of the logarithmic derivative of the transfer matrix (\ref{openT}), such that,
\bear
H&=&\frac{\sinh{\gamma}}{2}\frac{d}{d\lambda}\ln{t(\lambda)}\big|_{\lambda=0}  + const, \label{Hamiltonian} \\
 &=&\sum_{k=1}^{L-1} U_{k,k+1} + \frac{\sinh{\gamma}}{2} \frac{d K_1^{(-)}(\lambda)}{d\lambda}\Big|_{\lambda=0} +\frac{\tr_{\cal A}{\left[K_{\cal A}^{(+)}(0) U_{L,{\cal A}}\right]} }{\tr_{\cal A}{\left[K_{\cal A}^{(+)}(0)\right]}}. \nonumber
\ear

The solutions of the reflection equation (\ref{refle}) associated to the $R$-matrix (\ref{Rmatrix}) were recently obtained \cite{LIMA3}. Here we list only the diagonal ones, which we will use throughout this work.
\bear
K^{(-)}_{(1,0,0)}(\lambda)&=&\left(\begin{array}{ccc}
       k^{[I]}_{11}(\lambda) & 0 & 0 \\
       0 & 1 & 0 \\
       0 & 0 & 1 
\end{array} \right), \qquad
K^{(-)}_{(0,1,0)}(\lambda)=\left(\begin{array}{ccc}
       1 & 0 & 0 \\
       0 & k^{[I]}_{22}(\lambda) & 0 \\
       0 & 0 & 1 
\end{array} \right), \nonumber \\
K^{(-)}_{(0,0,1)}(\lambda)&=&\left(\begin{array}{ccc}
       1 & 0 & 0 \\
       0 & 1 & 0 \\
       0 & 0 & k^{[I]}_{33}(\lambda) 
\end{array} \right), 
\ear
where
\bear
k^{[I]}_{1,1}(\lambda) &=&\frac{\beta x_{2}(\lambda)\left[ (1+q^{2})x_{2}(\lambda)+x_{1}(\lambda) \right] +2\left[ x_{1}(\lambda)x_{2}^{\prime }(\lambda)-x_{1}^{\prime }(\lambda)x_{2}(\lambda)\right] }{-\beta x_{2}(\lambda)\left[ q^{-2}x_{2}(\lambda)+x_{1}(\lambda)\right] +2\left[ x_{1}(\lambda)x_{2}^{\prime }(\lambda)-x_{1}^{\prime }(\lambda)x_{2}(\lambda)\right] },  \nonumber\\ 
k^{[I]}_{2,2}(\lambda) &=&\frac{\beta x_{2}(\lambda)\left[ (q^{-2}+q^{2})x_{2}(\lambda)+x_{1}(\lambda)\right] +2\left[ x_{1}(\lambda)x_{2}^{\prime }(\lambda)-x_{1}^{\prime }(\lambda)x_{2}(\lambda)\right] }{-\beta x_{2}(\lambda)\left[ x_{2}(\lambda)+x_{1}(\lambda)\right] +2\left[ x_{1}(\lambda)x_{2}^{\prime }(\lambda)-x_{1}^{\prime }(\lambda)x_{2}(\lambda)\right] }, \nonumber \\ 
k^{[I]}_{3,3}(\lambda) &=&\frac{\beta x_{2}(\lambda)\left[ (q^{-2}+1)x_{2}(\lambda)+x_{1}(\lambda) \right] +2\left[ x_{1}(\lambda)x_{2}^{\prime }(\lambda)-x_{1}^{\prime }(\lambda)x_{2}(\lambda)\right] }{-\beta x_{2}(\lambda)\left[ q^{2}x_{2}(\lambda)+x_{1}(\lambda)\right] +2\left[ x_{1}(\lambda)x_{2}^{\prime }(\lambda)-x_{1}^{\prime }(\lambda)x_{2}(\lambda)\right]}, 
\ear
and 
\eq
x_1(\lambda)=\frac{\sinh(\gamma-\lambda)}{\sinh{\gamma}}, x_2(\lambda)=\frac{\sinh{\lambda}}{\sinh{\gamma}}. 
\en

There is an additional set of diagonal solutions of different form, given by
\bear
K^{(-)}_{(0,1,1)}(\lambda) &=&\left( \begin{array}{ccc} 1 & 0 & 0 \\ 0 & k^{[II]}_{1,1}(\lambda) & 0 \\ 0 & 0 & k_{1,1}^{[II]}(\lambda) \end{array} \right),
\quad 
K^{(-)}_{(1,0,1)}(\lambda)=\left( \begin{array}{ccc} k^{[II]}_{2,2}(\lambda) & 0 & 0 \\ 0 & 1 & 0 \\ 0 & 0 & k^{[II]}_{2,2}(\lambda)  \end{array} \right) , \nonumber \\
K^{(-)}_{(1,1,0)}(\lambda) &=&\left( \begin{array}{ccc} k^{[II]}_{3,3}(u) & 0 & 0 \\ 0 & k^{[II]}_{3,3}(\lambda) & 0 \\ 0 & 0 & 1 \end{array} \right) ,
\ear
where
\bear
k_{1,1}^{[II]}(\lambda) &=&\frac{\beta x_{2}(\lambda)\left[ q^{-2}x_{2}(\lambda)+x_{1}(\lambda) \right] +2\left[ x_{1}(\lambda)x_{2}^{\prime }(\lambda)-x_{1}^{\prime }(\lambda)x_{2}(\lambda)\right] }{-\beta x_{2}(\lambda)\left[ (1+q^{2})x_{2}(\lambda)+x_{1}(\lambda)\right] +2\left[ x_{1}(\lambda)x_{2}^{\prime }(\lambda)-x_{1}^{\prime }(\lambda)x_{2}(\lambda)\right] }, \nonumber \\ 
k^{[II]}_{2,2}(\lambda) &=&\frac{\beta x_{2}(\lambda)\left[ x_{2}(\lambda)+x_{1}(\lambda) \right] +2\left[ x_{1}(\lambda)x_{2}^{\prime }(\lambda)-x_{1}^{\prime }(\lambda)x_{2}(\lambda)\right] }{ -\beta x_{2}(\lambda)\left[ (q^{-2}+q^{2})x_{2}(\lambda)+x_{1}(\lambda)\right] +2 \left[ x_{1}(\lambda)x_{2}^{\prime }(\lambda)-x_{1}^{\prime }(\lambda)x_{2}(\lambda)\right] }, \nonumber \\ 
k^{[II]}_{3,3}(\lambda) &=&\frac{\beta x_{2}(\lambda)\left[ q^{2}x_{2}(\lambda)+x_{1}(\lambda)\right] +2\left[ x_{1}(\lambda)x_{2}^{\prime }(\lambda)-x_{1}^{\prime }(\lambda)x_{2}(\lambda)\right] }{-\beta x_{2}(\lambda)\left[ (q^{-2}+1)x_{2}(\lambda)+x_{1}(\lambda)\right] +2\left[ x_{1}(\lambda)x_{2}^{\prime }(\lambda)-x_{1}^{\prime }(\lambda)x_{2}(\lambda)\right] }.
\ear

The boundary terms of Hamiltonian (\ref{Hamiltonian}) are directly obtained from the above $K^{(\pm)}$ matrices. In particular, the left boundary acting non-trivially in the site 1 has the form
\eq
B_1=\frac{\sinh \gamma }{2}\left. \frac{dK_{1}^{(-)}(\lambda )}{d\lambda }
\right\vert _{\lambda =0}=\left( 
\begin{array}{ccc}
l_{11} & 0 & 0 \\ 
0 & l_{22} & 0 \\ 
0 & 0 & l_{33}
\end{array}
\right)_1,
\label{left}
\en
while the right boundary acting in the last site $L$ is given by
\eq
B_L=\frac{\mathrm{Tr}_{\cal A}\left[K_{\cal A}^{(+)}(0)U_{L,{\cal A}}\right]}{\mathrm{Tr}_{A}\left[K_{A}^{(+)}(0)\right]}=
\left( \begin{array}{ccc}
r_{11} & 0 & 0 \\ 
0 & r_{22} & 0 \\ 
0 & 0 & r_{33}
\end{array}
\right)_L,
\en
where
\bear
r_{11} &=&\frac{q^{-2}k_{33}^{+}(0)}{k_{11}^{+}(0)+k_{22}^{+}(0)+k_{33}^{+}(0)},\quad r_{22}=\frac{k_{22}^{+}(0)}{%
k_{11}^{+}(0)+k_{22}^{+}(0)+k_{33}^{+}(0)},  \nonumber \\
r_{33} &=&\frac{q^{2}k_{11}^{+}(0)}{k_{11}^{+}(0)+k_{22}^{+}(0)+k_{33}^{+}(0)%
}.  \label{right}
\ear

Therefore, we can compute the left boundary terms (\ref{left}) as
\bear
B_{1}^{(1,[I])} &=&\left( 
\begin{array}{ccc}
\frac{\beta\sinh \gamma }{2}  & 0 & 0 \\ 
0 & 0 & 0 \\ 
0 & 0 & 0%
\end{array}%
\right)_1,\quad 
B_{1}^{(2,[I])}=\left( 
\begin{array}{ccc}
0 & 0 & 0 \\ 
0 & \frac{\beta\sinh \gamma }{2}  & 0 \\ 
0 & 0 & 0%
\end{array}\right)_1, \nonumber  \\
B_{1}^{(3,[I])} &=&\left( 
\begin{array}{ccc}
0 & 0 & 0 \\ 
0 & 0 & 0 \\ 
0 & 0 & \frac{\beta\sinh \gamma }{2}
\end{array}
\right)_1
\ear
corresponding to the left $K$-matrices $K_{(1,0,0)}^{(-)},K_{(0,1,0)}^{(-)}$
and $K_{(0,0,1)}^{(-)}$, respectively. 

Similarly, we have three more left boundaries corresponding to the $K$-matrices $
K_{(0,1,1)}^{(-)},K_{(1,0,1)}^{(-)}$ and $K_{(1,1,0)}^{(-)}$
\bear
B_{1}^{(1,[II])} &=&\left( 
\begin{array}{ccc}
0 & 0 & 0  \\ 
0 & \frac{\beta\sinh \gamma }{2}  & 0 \\ 
0 & 0 & \frac{\beta\sinh \gamma }{2}
\end{array}
\right)_L,\quad 
B_{1}^{(2,[II])}=\left( 
\begin{array}{ccc}
\frac{\beta\sinh \gamma }{2}  & 0 & 0 \\ 
0 & 0 & 0 \\ 
0 & 0 & \frac{\beta\sinh \gamma }{2}
\end{array}
\right)_L, \nonumber \\
B_{1}^{(3,[II])} &=&\left( 
\begin{array}{ccc}
\frac{\beta\sinh \gamma }{2}  & 0 & 0 \\ 
0 & \frac{\beta\sinh \gamma }{2}  & 0 \\ 
0 & 0 & 0%
\end{array}%
\right)_L,
\ear
where $\beta$ is the left boundary free parameter.

From the isomorphism (\ref{isomorphism}), we have the $K^{(+)}(\lambda )$-matrices evaluated at $\lambda =0$ given as,
\bear
K_{(1,0,0)}^{(+)}(0) &=&\left( 
\begin{array}{ccc}
q^{-2}j_{11}^{[I]} & 0 & 0 \\ 
0 & 1 &  0 \\ 
0 & 0 & q^{2}%
\end{array}%
\right) ,\qquad K_{(0,1,0)}^{(+)}(0)=\left( 
\begin{array}{ccc}
q^{-2} & 0 & 0 \\ 
0 & j_{22}^{[I]} & 0 \\ 
0 & 0 & q^{2}%
\end{array}%
\right) , \nonumber \\
K_{(0,0,1)}^{(+)}(0) &=&\left( 
\begin{array}{ccc}
q^{-2} & 0 & 0 \\ 
0 & 1 & 0 \\ 
0 & 0 & q^2 j_{33}^{[I]}
\end{array}\right),
\ear
where 
\bear
j_{11}^{[I]} &=&\frac{2+(1+q^{2})\alpha \sinh \gamma}{2  -q^{-2}\alpha \sinh \gamma},\qquad 
j_{22}^{[I]}=\frac{2+(q^{-2}+q^{2})\alpha \sinh \gamma}{2-\alpha \sinh \gamma}, \nonumber\\
j_{33}^{[I]} &=&\frac{2+(1+q^{-2})\alpha \sinh \gamma}{2 -q^2\alpha\sinh \gamma},
\ear
and
\bear
K_{(0,1,1)}^{(+)}(0) &=&\left( 
\begin{array}{ccc}
q^{-2} & 0 & 0 \\ 
0 & j_{11}^{[II]} &  0 \\ 
0 &  0 & q^{2}j_{11}^{[II]} \end{array}\right),\qquad 
K_{(1,0,1)}^{(+)}(0)=\left( 
\begin{array}{ccc}
q^{-2}j_{22}^{[II]} & 0 & 0 \\ 
0 & 1 & 0  \\ 
0 & 0 & q^{2}j_{22}^{[II]}
\end{array} \right) , \nonumber  \\
K_{(1,1,0)}^{(+)}(0) &=&\left( 
\begin{array}{ccc}
q^{-2}j_{33}^{[II]} & 0 & 0 \\ 
0 & j_{33}^{[II]} & 0 \\ 
0 & 0 & q^{2} \end{array}\right), 
\ear
where
\bear
j_{11}^{[II]} &=&\frac{2+q^{-2}\alpha \sinh \gamma}{2-(1+q^{2})\alpha  \sinh\gamma},
\quad 
j_{22}^{[II]}=\frac{2+\alpha  \sinh \gamma }{2-(q^{2}+q^{-2})\alpha \sinh \gamma}, \nonumber \\
j_{33}^{[II]} &=&\frac{2+q^{2}\alpha \sinh \gamma}{2-(1+q^{-2})\alpha  \sinh \gamma},
\ear
and $\alpha$ is the right boundary free parameter.

Therefore, using (\ref{right}) we can write the corresponding right boundary terms
\bear
B_L^{(1,[I])}&=& \left( 
\begin{array}{ccc}
\frac{2-q^{-2}\alpha \sinh \gamma}{4\cosh \gamma } & 0 & 0 \\ 
0 & \frac{2-q^{-2}\alpha \sinh \gamma}{4\cosh \gamma } & 0 \\ 
0 & 0 & \frac{2+(1+q^{2})\alpha \sinh \gamma}{4\cosh \gamma }
\end{array}\right)_L,  \nonumber\\
B_L^{(2,[I])}&=&  \left( 
\begin{array}{ccc}
\frac{2-\alpha \sinh \gamma}{4\cosh \gamma } & 0 & 0 \\ 
0 & \frac{2+(q^{2}+q^{-2})\alpha \sinh \gamma}{4\cosh \gamma } & 0 \\ 
0 & 0 & \frac{2-\alpha \sinh \gamma}{4\cosh \gamma }
\end{array}\right)_L, \\
B_L^{(3,[I])}&=&  \left( 
\begin{array}{ccc}
\frac{2+(1+q^{-2})\alpha\sinh \gamma}{4\cosh \gamma } & 0 & 0 \\ 
0 & \frac{2-q^{2}\alpha \sinh \gamma}{4\cosh \gamma } & 0 \\ 
0 & 0 & \frac{2-q^{2}\alpha \sinh \gamma}{4\cosh \gamma }
\end{array}\right)_L, \nonumber
\ear
and
\bear
B_{L}^{(1,[II])}&=&\left(
\begin{array}{ccc}
\frac{2+q^{-2}\alpha \sinh \gamma}{4\cosh \gamma} & 0  & 0 \\ 
0 & \frac{2+q^{-2}\alpha \sinh \gamma}{4\cosh \gamma} & 0 \\ 
0 & 0 & \frac{2-(1+q^2)\alpha \sinh \gamma}{4\cosh \gamma}
\end{array}\right), \nonumber \\
B_{L}^{(2,[II])}&=&\left( 
\begin{array}{ccc}
\frac{2+\alpha \sinh \gamma}{4\cosh \gamma} & 0 & 0 \\ 
0 & \frac{2-(q^2+q^{-2})\alpha \sinh \gamma}{4\cosh \gamma} & 0 \\ 
0 & 0 & \frac{2+\alpha \sinh \gamma}{4\cosh \gamma} \end{array} \right), \\
B_{L}^{(3,[II])}&=&\left(
\begin{array}{ccc}
\frac{2-(1+q^{-2})\alpha \sinh \gamma}{4\cosh \gamma} & 0  & 0 \\ 
0 & \frac{2+q^{2}\alpha \sinh \gamma}{4\cosh \gamma} & 0 \\ 
0 & 0 & \frac{2+q^{2}\alpha \sinh \gamma}{4\cosh \gamma}
\end{array}\right)_L. \nonumber
\ear

Here we have $6$ different integrable boundaries related by the isomorphism (\ref{isomorphism}). However, it is worth to note that other combination of the boundaries are allowed $B_{1,L}^{(i,j,[a,b])}=B_1^{(i,[a])}+B_L^{(j,[b])}$ with $i,j=1,2,3$, $a,b=I,II$ resulting in $36$ integrable boundaries for the spin-$1$ $U_{q}[sl(2)]$ Temperley-Lieb Hamiltonian.

The action of the boundary terms on the Hilbert space is given by
\eq
B_{1,L}^{(i,j,[a,b])}\ket{\overset{1}{\sigma}\cdots \overset{L}{\tau}}={\cal E} _{\sigma \tau}^{(i,j,[a,b])}|\overset{1}{\sigma}\cdots \overset{L}{\tau}>
\en
where ${\cal E}_{\sigma \tau}^{(i,j,[a,b])}=l_{\sigma \sigma}^{(i,[a])}+r_{\tau \tau}^{(j,[b])}$ and the sites are indexed by $\sigma,\tau=(1,2,3)\doteq (+,0,-)$. Here we recall that $l_{\sigma \sigma}^{(i,[a])}$ and $r_{\tau \tau}^{(j,[b])}$ are the matrix elements of the boundary matrices $B_1^{(i,[a])}$ and $B_L^{(j,[b])}$ respectively.

In the next section, we will restrict ourselves to the case of integrable boundaries related by the isomorphism ($B_{1,L}^{(i,[a])}=B_1^{(i,[a])}+B_L^{(i,[a])}$) and we shall use the coordinate Bethe ansatz in order to obtain the eigenvalues of the Hamiltonian (\ref{Hamiltonian}).

\section{Coordinate Bethe ansatz}
\label{cBA}

In most of the cases where Bethe ansatz is successfully applied, one can build up all the eigenstates from just one reference state and usually there exist only few of such reference states available. In the case of Temperley-Lieb spin chains, by contrast, we have exponentially degenerated ground states, which implies that we have a very large number of reference states. In fact we have $3 \times 2^{L-1}$ natural eigenstates which can be used as reference states. This explains the difficulties in constructing all the eigenstates from just one reference state \cite{LIMA1,LIMA2,GHIOTTO}.

\begin{table}[h]
\begin{center}
\begin{tabular}{|l|l|l|}
\hline
${\cal E}_{11}
\begin{cases}
\ket{++++} \\
\ket{+0++} \\
\ket{++0+}
\end{cases}
$
 & ${\cal E}_{12}
\begin{cases}
\ket{+++0} \\
\ket{+0+0} \\
\ket{+0-0}
\end{cases}
$
& ${\cal E}_{13}
\begin{cases}
\ket{++0-} \\
\ket{+0--} 
\end{cases} $ \\ 
\hline
${\cal E}_{21}
\begin{cases}
\ket{0+++} \\
\ket{0+0+} \\
\ket{0-0+}
\end{cases}
$
 & ${\cal E}_{22}
\begin{cases}
\ket{0++0} \\
\ket{0--0} 
\end{cases}
$
& ${\cal E}_{23}
\begin{cases}
\ket{0---} \\
\ket{0-0-} \\
\ket{0+0-} 
\end{cases} $ \\ 
\hline
${\cal E}_{31}
\begin{cases}
\ket{-0++} \\
\ket{--0+} 
\end{cases}
$
 & ${\cal E}_{32}
\begin{cases}
\ket{---0} \\
\ket{-0-0} \\
\ket{-0+0}  
\end{cases}
$
& ${\cal E}_{33}
\begin{cases}
\ket{----} \\
\ket{--0-} \\
\ket{-0--} 
\end{cases} $ \\ 
\hline
\end{tabular}
\end{center}
\caption{The reference states (natural eigenstates) of the Hamiltonian (\ref{Hamiltonian}) and its eigenvalues for $L=4$.}
\label{tab1}
\end{table}

The reason for such differences is that the bulk part of the Hamiltonian (the Temperley-Lieb operator $U_{k,k+1}$) is the projector operator onto the two-site spin zero singlet. This implies that there exist $3 \times 2^{L-1}$ states which are eigenstates of the bulk Hamiltonian with zero eigenvalues. Therefore, these states are also eigenstates of the boundary part of the Hamiltonian $B_{1,L}^{(i,[a])}$ with eigenvalues ${\cal E}_{\sigma \tau}^{(i,[a])}$ (see e.g Table \ref{tab1}).

Moreover, apart from the natural degenerescence of the boundary eigenvalues ${\cal E}_{\sigma \tau}^{(i,[a])}$, one can see from the structure of the boundary matrix $K^{(\pm)}$ that not all ${\cal E}_{\sigma \tau}^{(i,[a])}$ are independent. In fact they are also degenerated and can be grouped in four blocks for each integrable boundary related by the isomorphism (see Table \ref{tab2}). 

\begin{table}
\begin{center}
\begin{tabular}{|c|c|}
\hline
${\cal E}_{11}^{(1,[a])}={\cal E}_{12}^{(1,[a])}$ & ${\cal E}_{13}^{(1,[a])}$ \\ 
\hline
${\cal E}_{21}^{(1,[a])}={\cal E}_{22}^{(1,[a])}={\cal E}_{31}^{(1,[a])}={\cal E}_{32}^{(1,[a])}$ & ${\cal E}_{33}^{(1,[a])}={\cal E}_{23}^{(1,[a])}$ \\
\hline
\hline
${\cal E}_{12}^{(2,[a])}={\cal E}_{32}^{(2,[a])}$ & ${\cal E}_{22}^{(2,[a])}$ \\ 
\hline
${\cal E}_{11}^{(2,[a])}={\cal E}_{13}^{(2,[a])}={\cal E}_{31}^{(2,[a])}={\cal E}_{33}^{(2,[a])}$ & ${\cal E}_{21}^{(2,[a])}={\cal E}_{23}^{(2,[a])}$ \\
\hline\hline
${\cal E}_{11}^{(3,[a])}={\cal E}_{21}^{(3,[a])}$ & ${\cal E}_{31}^{(3,[a])}$ \\ 
\hline
 ${\cal E}_{12}^{(3,[a])}={\cal E}_{13}^{(3,[a])}={\cal E}_{22}^{(3,[a])}={\cal E}_{23}^{(3,[a])}$ & ${\cal E}_{32}^{(3,[a])}={\cal E}_{33}^{(3,[a])}$ \\
\hline
\end{tabular}
\end{center}
\caption{The relation among the boundary eigenvalues for different solution of the reflection equation. This relations hold true for any $a=I,II$.}
\label{tab2}
\end{table}

In face of the large number of reference states, the standard construction of the all eigenstates seems to be impracticable, though it is possible. However, in order to obtain the eigenvalues of the Hamiltonian it is enough to work out with a few reference states. In fact, we can take one reference state from each block of eigenvalues ${\cal E}_{\sigma \tau}^{(i,[a])}$. From now on, we drop the label for different solutions of the reflection equation from the boundary eigenvalues, such that ${\cal E}_{\sigma \tau}^{(i,[a])}={\cal E}_{\sigma \tau}$.

\subsection{Ferromagnetic reference state}

We shall start by considering the pseudo particle as a singlet over the standard ferromagnetic state. Therefore, it is convenient to start our ansatz with the following linear combination of the basis states \cite{GHIOTTO},
\eq
\ket{\Omega(k)}=q^{-2} \ket{+\cdots \overset{k}{+} - +\cdots +} -q^{-1} \ket{+\cdots +\overset{k}{0}0 +\cdots +} +  \ket{+\cdots + \overset{k}{-}+ \cdots +},
\en
which is an eigenstate of $U_{k,k+1}$ such that
\begin{align}
&U_{k,k+1} \ket{\Omega(k)} = Q\ket{\Omega(k)}, & U_{k+ 1,k+2} \ket{\Omega(k)} &= \ket{\Omega(k+1)}, \nonumber\\
&U_{k,k+1} \ket{\Omega(k\pm 1)} = \ket{\Omega(k)}, & U_{k- 1,k} \ket{\Omega(k)} &= \ket{\Omega(k-1)}, \\
&U_{k,k+1} \ket{\Omega(j)} = 0, ~ \mbox{if } k\neq \{j,j+1\},  \nonumber
\end{align}
where $Q=(q^2+1+q^{-2})$. Therefore, the action of the Hamiltonian $H=\sum_{k=1}^{L-1} U_{k,k+1}+B_{1,L}$ over this state results,
\bear
H\ket{\Omega(k)}=Q \ket{\Omega(k)}+ {\cal E}_{11}\ket{\Omega(k)} + \ket{\Omega(k-1)} + \ket{\Omega(k+1)},  1<k<L-1 
\ear
\bear
H\ket{\Omega(1)}&=&Q \ket{\Omega(1)}+ {\cal E}_{11}\ket{\Omega(1)} + \ket{\Omega(0)} + \ket{\Omega(2)} \\
H\ket{\Omega(L-1)}&=&Q \ket{\Omega(L-1)}+ {\cal E}_{11}\ket{\Omega(L-1)} + \ket{\Omega(L-2)} + \ket{\Omega(L)},
\ear
where $B_{1,L} \ket{\Omega(k)}={\cal E}_{11}\ket{\Omega(k)}, 1<k<L-1$, $\ket{\Omega(0)}=(B_{1,L}-{\cal E}_{11})\ket{\Omega(1)}$ and $\ket{\Omega(L)}=(B_{1,L}-{\cal E}_{11})\ket{\Omega(L-1)}$. In addition to the previous relation, we have a set of closing relations
\bear
H\ket{\Omega(0)}&=&\Delta_l^{(1)} \ket{\Omega(1)}+{\cal E}_{v_{i,1} 1}  \ket{\Omega(0)}, \\
H\ket{\Omega(L)}&=&\Delta_r^{(1)} \ket{\Omega(L-1)}+ {\cal E}_{1 u_{i,1}} \ket{\Omega(L)},
\ear
where $\Delta_l^{(1)}=({\cal E}_{21}-{\cal E}_{11})+q^2({\cal E}_{31}-{\cal E}_{11}) $, $\Delta_r^{(1)}=({\cal E}_{12}-{\cal E}_{11})+q^{-2}({\cal E}_{13}-{\cal E}_{11})$. In order to cover all the solution of the reflection equation, we introduce the following notation $\vec{v}_{1}=(2,2,3)$ and $\vec{u}_{1}=(3,2,2)$ whose elements $i=1,2,3$ represent different solution of the reflection equation (for any $a=I,II$). In the above relations, we exploited the fact that,
\bear
U_{1,2}\ket{\Omega(0)}&=&\Delta_l^{(1)} \ket{\Omega(1)}, \\
B_{1,L}\ket{\Omega(0)}&=&{\cal E}_{v_{i,1} 1}  \ket{\Omega(0)}, \\
U_{L-1,L}\ket{\Omega(L)}&=&\Delta_r^{(1)} \ket{\Omega(L-1)}, \\
B_{1,L}\ket{\Omega(L)}&=&{\cal E}_{1 u_{i,1}}  \ket{\Omega(L)}.
\ear

\subsubsection{One-particle state}

In the first non-trivial sector, we assume the following ansatz for the eigenstates
\eq
\ket{\Psi_1}=\sum_{k=1}^{L-1} A(k)\ket{\Omega(k)}.
\en

Imposing the eigenvalue equation $H\ket{\Psi_1}=E_1\ket{\Psi_1}$ is fulfilled, we obtain a set of equations for the function $A(k)$,
\bear
(Q+{\cal E}_{11}-E_1)A(k)+A(k-1)+A(k+1)&=&0, ~ 1<k<L-1 \label{eqk} \\
({\cal E}_{v_{i,1} 1}-E_1)A(0)+\Delta_l^{(1)} A(1)&=&0, \label{eq1}\\
({\cal E}_{1u_{i,1}}-E_1)A(L)+\Delta_r^{(1)} A(L-1)&=&0. \label{eqL}
\ear
Taking the ansatz for the amplitude
\eq
A(k)=a(\theta)\xi^k -a(-\theta)\xi^{-k},
\en
and substituting in Eq.(\ref{eqk}) provides the following expression for the energy eigenvalues
\eq
E_1={\cal E}_{11}+ Q+\xi + \xi^{-1}.
\en
The parameter $\xi$ and the ratio of the amplitudes $a(\theta)/a(-\theta)$ are fixed by the equations (\ref{eq1}-\ref{eqL}), which results in the Bethe ansatz equation
\eq
\xi^{2 L}=\left( \frac{Q+{\cal E}_{11}-{\cal E}_{v_{i,1} 1}+ \xi^{-1} + \xi(1-\Delta_l^{(1)})}{Q+{\cal E}_{11}-{\cal E}_{v_{i,1} 1}+ \xi + \xi^{-1}(1-\Delta_l^{(1)})} \right)\left( \frac{Q +{\cal E}_{11}-{\cal E}_{1u_{i,1}}+ \xi^{-1} + \xi(1-\Delta_r^{(1)})}{Q +{\cal E}_{11}-{\cal E}_{1u_{i,1}}+ \xi + \xi^{-1}(1-\Delta_r^{(1)})} \right).
\en

\subsubsection{Two-particle state}

In the next particle sector, we have two interacting pseudo-particles, which can be represented as a product of two pseudo-particles eigenstates, as given by
\eq
\ket{\Psi_2}=\sum_{k_1+1<k_2} A(k_1,k_2)\ket{\Omega(k_1,k_2)},
\en
where
\eq
\ket{\Omega(k_1,k_2)}=\sum_{i,j=-1}^{1} (-1)^{i+j}q^{i+j-2} \ket{k_1(-i,i);k_2(-j,j)}.
\en

We can split the action of the Hamiltonian on the state $\ket{\Omega(k_1,k_2)}$ in four cases:

\noindent(i) The case where two pseudo-particles are separated in the bulk,
\bear
H\ket{\Omega(k_1,k_2)}=(2 Q+{\cal E}_{11} )\ket{\Omega(k_1,k_2)}+\ket{\Omega(k_1-1,k_2)}+\ket{\Omega(k_1+1,k_2)} \\
+\ket{\Omega(k_1,k_2-1)}+\ket{\Omega(k_1,k_2+1)}, \qquad 1<k_1+2 <k_2<L-1 \nonumber 
\ear
(ii) The case where the pseudo-particles are separated but one of them or both are at the boundaries
\begin{align}
&H\ket{\Omega(1,k_2)}=(2 Q+{\cal E}_{11} )\ket{\Omega(1,k_2)}+\ket{\Omega(0,k_2)}+\ket{\Omega(2,k_2)} \nonumber\\
&+\ket{\Omega(1,k_2-1)}+\ket{\Omega(1,k_2+1)}, \qquad 5< k_2 <L-1  \\
&H\ket{\Omega(k_1,L-1)}=(2 Q+{\cal E}_{11} )\ket{\Omega(k_1,L-1)}+\ket{\Omega(k_1-1,L-1)} \nonumber  \\
&+\ket{\Omega(k_1+1,L-1)}+\ket{\Omega(k_1,L-1)}+\ket{\Omega(k_1,L)}, ~ 1<k_1<L-3 \\
&H\ket{\Omega(1,L-1)}=(2 Q+{\cal E}_{11} )\ket{\Omega(1,L-1)}+\ket{\Omega(0,L-1)}+\ket{\Omega(2,L-1)} \nonumber\\
&+\ket{\Omega(1,L-1)}+\ket{\Omega(1,L)},  
\end{align}
(iii) The case where the particles are neighbours in the bulk
\begin{align}
&H\ket{\Omega(k,k+2)}=(2 Q+{\cal E}_{11} )\ket{\Omega(k,k+2)}+\ket{\Omega(k-1,k+2)}+\ket{\Omega(k,k+3)}\nonumber\\
&+\ket{\Omega(k+1,k+2)} +\ket{\Omega(k,k+1)}, \qquad 1<k<L-3 
\end{align}
(iv) The case where the particles are neighbours at the boundaries
\begin{align}
&H\ket{\Omega(1,3)}=(2 Q+{\cal E}_{11} )\ket{\Omega(1,3)}+\ket{\Omega(0,3)}+\ket{\Omega(2,3)}  \nonumber\\
&+\ket{\Omega(1,2)}+\ket{\Omega(1,4)},  \\
&H\ket{\Omega(L-3,L-1)}=(2 Q+{\cal E}_{11} )\ket{\Omega(L-3,L-1)}+\ket{\Omega(L-4,L-1)}   \nonumber\\
&+\ket{\Omega(L-2,L-1)}+\ket{\Omega(L-3,L-2)}+\ket{\Omega(L-3,L)}.
\end{align}

In the above relations, we have introduced new states whose definition are given by
\begin{align}
\ket{\Omega(0,k_2)}&=(B_{1,L}-{\cal E}_{11}) \ket{\Omega(1,k_2)}, \\
\ket{\Omega(k_1,L)}&=(B_{1,L}-{\cal E}_{11}) \ket{\Omega(k_1,L-1)}, \\
\ket{\Omega(0,L-1)}+\ket{\Omega(1,L)} &= (B_{1,L}-{\cal E}_{11}) \ket{\Omega(1,L-1)},\\
\ket{\Omega(k+1,k+2)}+\ket{\Omega(k,k+1)}&= U_{k+1,k+2} \ket{\Omega(k,k+2)} .
\end{align}
The action of the Hamiltonian on these states can be written as follows
\begin{align}
H\ket{\Omega(0,k_2)}&=\Delta_l^{(1)} \ket{\Omega(1,k_2)} + {\cal E}_{v_{i,1} 1}\ket{\Omega(0,k_2)} +\ket{\Omega(0,k_2-1)} \nonumber \\
&+ Q\ket{\Omega(0,k_2)} +\ket{\Omega(0,k_2+1)}, \\
H\ket{\Omega(k_1,L)}&=\Delta_r^{(1)} \ket{\Omega(k_1,L-1)} + {\cal E}_{1u_{i,1}}\ket{\Omega(k_1,L)} +\ket{\Omega(k_1-1,L)} \nonumber \\
&+ Q\ket{\Omega(k_1,L)} +\ket{\Omega(k_1+1,L)}, \\
H\ket{\Omega(k,k+1)}&=Q \ket{\Omega(k,k+1)} + \ket{\Omega(k-1,k+1)} +\ket{\Omega(k,k+2)}  \nonumber \\
& + {\cal E}_{11}\ket{\Omega(k,k+1)}.
\end{align}

In order to obtain the eigenvalues, we have to substitute the above relations in the eigenvalue equation ($H\ket{\Psi_2}=E_2\ket{\Psi_2}$). This will provide us the following set of equations for the amplitude $A(k_1,k_2)$,
\begin{align}
&(2 Q+{\cal E}_{11} -E_2)A(k_1,k_2)+A(k_1-1,k_2)+A(k_1+1,k_2)+A(k_1,k_2-1) \nonumber\\
&+A(k_1,k_2+1)=0, \label{bulk}\\
&(Q+{\cal E}_{11} -E_2)A(k,k+1)+A(k-1,k+1)+A(k,k+2)=0, \label{near}
\end{align}
\begin{align}
&(Q+{\cal E}_{v_{i,1} 1} -E_2)A(0,k_2)+A(0,k_2-1)+A(0,k_2+1)+ \Delta_l^{(1)} A(1,k_2)=0, \label{btl}\\
&(Q+{\cal E}_{1u_{i,1}} -E_2)A(k_1,L)+A(k_1-1,L)+A(k_1+1,L)+\Delta_r^{(1)} A(k_1,L-1)=0. \label{btr}
\end{align}

One can obtain the eigenvalues from the equation (\ref{bulk}),
\eq
E_2={\cal E}_{11} + 2Q+ \xi_1 +\xi_1^{-1}+ \xi_2 +\xi_2^{-1},
\label{E2}
\en
provided that the following parametrization for the amplitudes is assumed
\eq
A(k_1,k_2)=\sum_{P}\varepsilon_{P}a(\theta_1,\theta_2)\xi_1^{k_1}\xi_2^{k_2},
\label{BA2}
\en
where the sum extends over all permutations and negations of momenta ($\theta_i$), such that $\xi_i=e^{\im \theta_i}$, and $\varepsilon_P$ is the signature of permutations and negations. This structure already reflects the existence of the boundary.

On the other hand, the equation (\ref{near}) is the meeting condition for the two pseudo-particle states. Using the ansatz (\ref{BA2}), we obtain the following phase shifts,
\begin{align}
a(\theta_2,\theta_1)&=\left(\frac{s(\theta_2,\theta_1)}{s(\theta_1,\theta_2)}\right)a(\theta_1,\theta_2), \label{phase1} \\ a(\theta_2,-\theta_1)&=\left(\frac{s(\theta_2,-\theta_1)}{s(-\theta_1,\theta_2)}\right)a(-\theta_1,\theta_2), \\
a(-\theta_2,\theta_1)&=\left(\frac{s(-\theta_2,\theta_1)}{s(\theta_1,-\theta_2)}\right)a(\theta_1,-\theta_2), \\
a(-\theta_2,-\theta_1)&=\left(\frac{s(-\theta_2,-\theta_1)}{s(-\theta_1,-\theta_2)}\right)a(-\theta_1,-\theta_2), \label{phase4}
\end{align}
where
\eq
s(\theta_1,\theta_2)=1+\xi_1 \xi_2 + \xi_1 Q.
\en

At this point, we still have two remaining equation (\ref{btl}-\ref{btr}) which introduce the boundary effects. One can introduce the expressions (\ref{E2}-\ref{BA2}) in the equation for the left boundary (\ref{btl}), which results
\begin{align}
a(-\theta_1,\theta_2)&=F_l(\theta_1)a(\theta_1,\theta_2),\\ 
a(-\theta_2,\theta_1)&=F_l(\theta_2)a(\theta_2,\theta_1), 
\end{align}
where
\bear
F_l(\theta_1)=\left(\frac{Q+{\cal E}_{11}-{\cal E}_{v_{i,1} 1} + \xi_1^{-1} +\xi_1 (1-\Delta_l^{(1)})}{Q+{\cal E}_{11}-{\cal E}_{v_{i,1} 1}+ \xi_1 +\xi_1^{-1} (1-\Delta_l^{(1)})}\right).
\ear
Likewise for right boundary, one obtain the the following relations
\begin{align}
a(\theta_2,-\theta_1)&=\xi_1^{2 L}F_r(\theta_1)a(\theta_2,\theta_1), \\ 
a(\theta_1,-\theta_2)&=\xi_2^{2 L}F_r(\theta_2)a(\theta_1,\theta_2), 
\end{align}
where
\bear
F_r(\theta_1)=\left(\frac{Q+{\cal E}_{11}-{\cal E}_{1u_{i,1}} + \xi_1 +\xi_1^{-1} (1-\Delta_r^{(1)})}{Q+{\cal E}_{11}-{\cal E}_{1u_{i,1}}+ \xi_1^{-1} +\xi_1 (1-\Delta_r^{(1)})}\right).
\ear

Combining these relations with the phase shift relations (\ref{phase1}-\ref{phase4}), we obtain the Bethe ansatz equations
\bear
\xi_1^{2 L}=F_{l}(\theta_1)F_{r}(\theta_1)^{-1} \left(\frac{s(\theta_1,\theta_2)}{s(\theta_2,\theta_1)}\right)\left(\frac{s(\theta_2,-\theta_1)}{s(-\theta_1,\theta_2)}\right), \\
\xi_2^{2 L}=F_{l}(\theta_2)F_{r}(\theta_2)^{-1} \left(\frac{s(\theta_2,\theta_1)}{s(\theta_1,\theta_2)}\right)\left(\frac{s(\theta_1,-\theta_2)}{s(-\theta_2,\theta_1)}\right).
\ear

\subsubsection{General sector}

The generalization to any number $n$ of pseudo-particles goes along the same lines as before, although the calculation becomes cumbersome. Therefore, we just present the final results. The energy eigenvalues are given by the sum of single pseudo-particle energies
\eq
E_n={\cal E}_{11} + \sum_{i=1}^n Q+\xi_i +\xi_i^{-1},
\label{En}
\en
and the corresponding Bethe ansatz equations depend on the phase shift of two pseudo-particles and on the boundary factors
\bear
\xi_i^{2 L}=F_{l}(\theta_i)F_{r}(\theta_i)^{-1} \prod_{\stackrel{j=1}{j\neq i}}^{n}\left(\frac{s(\theta_i,\theta_j)}{s(\theta_j,\theta_i)}\right)\left(\frac{s(\theta_j,-\theta_i)}{s(-\theta_i,\theta_j)}\right).
\label{BAn}
\ear

\subsection{Other reference states}

In order to obtain the whole spectrum of the Hamiltonian we have to consider additional reference states. This has to be done for each different boundary eigenvalues ${\cal E}_{\sigma\tau}$. As a result of that, we must have as many as reference states and consequently Bethe ansatz equations as boundary eigenvalues.

In principle, we have nine boundary eigenvalues ${\cal E}_{\sigma\tau}$. If one choose one reference state for each boundary eigenvalues (e.g the first state of each block of Table \ref{tab1} extended to $L$-sites) and proceed along the same lines as the previous section, we obtain nine eigenvalues expressions 
\eq
E_n^{(\sigma,\tau)}={\cal E}_{\sigma \tau } + \sum_{i=1}^n Q+\xi_i +\xi_i^{-1},
\label{Enr}
\en
as well as its associated Bethe ansatz equations
\bear
\xi_i^{2 L}=F_{l}^{(\sigma,\tau)}(\theta_i)F_{r}^{(\sigma,\tau)}(\theta_i)^{-1} \prod_{\stackrel{j=1}{j\neq i}}^{n}\left(\frac{s(\theta_i,\theta_j)}{s(\theta_j,\theta_i)}\right)\left(\frac{s(\theta_j,-\theta_i)}{s(-\theta_i,\theta_j)}\right), 
\label{BAnr}
\ear
where
\bear
F_l^{(\sigma,\tau)}(\theta_1)&=&\left(\frac{Q+{\cal E}_{\sigma\tau}-{\cal E}_{v_{i\sigma} \tau} + \xi_1 +\xi_1^{-1} (1-\Delta_l^{(\sigma)})}{Q+{\cal E}_{\sigma\tau}-{\cal E}_{v_{i\sigma} \tau} + \xi_1^{-1} +\xi_1 (1-\Delta_l^{(\sigma)})}\right), \\
F_r^{(\sigma,\tau)}(\theta_1)&=&\left(\frac{Q+{\cal E}_{\sigma\tau}-{\cal E}_{\sigma u_{i\tau}} + \xi_1 +\xi_1^{-1} (1-\Delta_r^{(\tau)})}{Q+{\cal E}_{\sigma\tau}-{\cal E}_{\sigma u_{i\tau}} + \xi_1^{-1} +\xi_1 (1-\Delta_r^{(\tau)})}\right),
\ear
\bear
\Delta_l^{(2)}&=&({\cal E}_{31}-{\cal E}_{21}) q^{2}+ ({\cal E}_{11}-{\cal E}_{21}) q^{-2},\\
\Delta_r^{(2)}&=&({\cal E}_{11}-{\cal E}_{12}) q^{2}+ ({\cal E}_{13}-{\cal E}_{12}) q^{-2}, \\
\Delta_l^{(3)}&=&({\cal E}_{21}-{\cal E}_{31}) + ({\cal E}_{11}-{\cal E}_{31}) q^{-2},\\
\Delta_r^{(3)}&=&({\cal E}_{11}-{\cal E}_{13}) q^{2}+ ({\cal E}_{12}-{\cal E}_{13}) .
\ear
The remaining index $v_{i,\sigma}$ are defined by $\vec{v}_{2}=(1,1,3)$, $\vec{v}_{3}=(1,1,2)$ and the $u_{i,\tau}$ are given by $\vec{u}_{2}=(3,1,1)$, $\vec{u}_{3}=(2,1,1)$.

However, we can see from Table \ref{tab2} that most of these equations degenerate into each other, resulting in four equations for each integrable boundary. We have verified numerically the completeness of the spectrum up to $L=6$ sites.

\section{Conclusion}
\label{CONCLUSION}

In this paper we obtained the spectrum of the spin-$1$ $U_q\left[sl(2)\right]$ Temperley-Lieb spin chain with diagonal open boundary conditions. We have identified that this model has large number of possible reference states. By selecting a small subset of these states, we manage to obtain four eigenvalue expressions and its associated Bethe ansatz equations by means of a generalization of the coordinate Bethe ansatz. This provides the complete description of the spectrum of the model for any values of the boundary parameters. We verified that the Bethe ansatz results are in agreement with the direct diagonalization of the Hamiltonian up to six sites. 

Apart from the news results, we believe that this work also brings a better understanding on the coordinate Bethe ansatz construction of the eigenstates. Although with this new perspective is possible to construct all the eigenstates for finite system size, this seems to be rather impracticable. Therefore, we still leave the problem of counting of the spectral multiplicities as an open question. We also hope that this work would shed some light on the algebraic Bethe ansatz construction for the Temperley-Lieb spin chains.

\section*{Acknowledgments}
GAPR thanks Andreas Kl\"umper for discussions. This work has been supported by FAPESP and CNPq.


\begin{thebibliography}{100}
\bibitem{BAXTER} R.J. Baxter, {\it Exactly Solved Models in Statistical Mechanics}, Academic Press, New York, 1982.
\bibitem{KOREPIN} V.E. Korepin, A.G. Izergin and N.M. Bogoliubov, {\it Quantum Inverse Scattering Method, Correlation Functions and Algebraic Bethe Ansatz}, Cambridge Univ. Press, Cambridge, 1992.
\bibitem{PARKI} J.B. Parkinson, J. Phys. C: Sol. State Phys. 21 (1988) 3793.
\bibitem{KLUMPER90} A. Kl\"umper, J. Phys. A: Math. Gen. 23 (1990) 809.
\bibitem{ALCARAZ} F.C. Alcaraz, R. K\"oberle and A. Lima-Santos, Int. J. Mod. Phys. A 7 (1992) 7615.
\bibitem{BATCHELOR3} M.N. Barber, M.T. Batchelor, Phys. Rev. B 40 (1989) 4621.
\bibitem{BATCHELOR1} M.T. Batchelor, L. Mezincescu, R.I. Nepomechie and V. Rittenberg, J. Phys. A: Math. Gen. 23 (1990) L141.
\bibitem{BATCHELOR2} M.T. Batchelor and A. Kuniba, J. Phys. A: Math. Gen. 24 (1991) 2599.
\bibitem{LIMA1} R. K\"oberle, A. Lima-Santos, J. Phys. A: Math. Gen. 27 (1994) 5409; J. Phys. A: Math. Gen. 29 (1996) 519.
\bibitem{LIMA2} A. Lima-Santos and R.C.T. Ghiotto, J. Phys. A: Math. Gen. 31 (1998) 505.
\bibitem{GHIOTTO} R.C.T. Ghiotto and A. Malvezzi, Int. J. Mod. Phys. A, 15 (2000) 3395.
\bibitem{KULISH1} P.P. Kulish, J. Phys. A: Math. Gen. 36 (2003) L489.
\bibitem{KLUMPER} B. Aufgebauer, A. Kl\"umper, J. Stat. Mech. (2010) P05018.
\bibitem{LIMA3} A. Lima-Santos, J. Stat. Mech. (2011) P01009.
\bibitem{KULISH2} J. Avan, P.P. Kulish and G. Rollet, Theor. Math. Phys. 169 (2011) 1530.
\bibitem{MARTIN} D. Levy and P. Martin, J. Phys. A: Math. Gen. 27 (1994) L521.
\bibitem{SKLYANIN} I. Cherednik, Theor. Math. Phys. 61 (1984) 977; E.K. Sklyanin, J. Phys. A: Math. Gen. 21 (1988) 2375.
\bibitem{NEPO} L. Mezincescu, R.I. Nepomechie, J. Phys. A: Math. Gen. 24 (1991) 217.
\end{thebibliography}
\end{document}